\documentclass[12pt]{iopart}

\begin{document}

\title[Boost invariance of the gravitational field dynamics]
{Boost invariance of the gravitational field dynamics: quantization without time gauge.}

\author{Francesco Cianfrani$^1$ and Giovanni Montani$^{123}$}

\address{$^1$ICRA-International Center for Relativistic Astrophysics\\
Dipartimento di Fisica (G9), Universit\`a  di Roma, ``La Sapienza",\\
Piazzale Aldo Moro 5, 00185 Rome, Italy.}

\vspace{0.5cm}

\address{$^2$ENEA C.R. Frascati (Dipartimento F.P.N.),\\ via Enrico Fermi 45, 00044 Frascati, Rome, Italy.}

\vspace{0.5cm}

\address{$^3$ICRANET C. C. Pescara, \\ Piazzale della Repubblica, 10, 65100 Pescara, Italy.}

\ead{francesco.cianfrani@icra.it, montani@icra.it}

\begin{abstract}
We perform a canonical quantization of gravity in a second-order formulation, taking as configuration variables those describing a 4-bein, not adapted to the space-time splitting. We outline how, neither if we fix the Lorentz frame before quantizing, nor if we perform no gauge fixing at all, is invariance under boost transformations affected by the quantization. 
\end{abstract}

\vspace{2pc}
\maketitle

\section{Introduction}

The development of a quantum theory for the gravitational field is one of the main points in Theoretical Physics. The most promising approaches in such direction are those of String Theory \cite{Pol} and of Loop Quantum Gravity (LQG) \cite{Rov}. While String Theory implies a completely new interpretation of all fields, as properties due to the vibration of fundamental strings, and it till now provides just a perturbative approach to Quantum Gravity, LQG is a more conservative attempt toward a non-perturbative canonical quantization of space-time geometry.\\ 
LQG is based on a reformulation of General Relativity in terms of $SU(2)$ connections (Barbero-Immirzi connections \cite{Ba95}), where the phase space is that of an $SU(2)$ gauge theory. This kind of reduction of the Lorentz group to a compact one is a key point of LQG, since it allows for the use of standard techniques of gauge theories (Wilson loop) in view of a canonical non-perturbative quantization. But, after the quantization, an ambiguity arises, in terms of the $\gamma$ parameter (Immirzi parameter) which enters the spectrum of physical observables. The physical interpretation of $\gamma$ is still under investigation. While standard works on LQG treat it as a fundamental parameter, fixed by the request of reproducing results on the entropy of Black Holes \cite{ABCK,Kr}, nevertheless there are authors who consider it as an ambiguity due to the breaking of some symmetry \cite{GM,A00}. In particular, the debate is on the fate of Lorentz invariance in the Barbero-Immirzi formulation. This formulation is based on fixing, before quantizing, the so-called time-gauge condition, which corresponds to set the 4-bein vectors, such that the time-like one $e_0$ is normal to spatial hypersurfaces. If this hypothesis is neglected, a deep complication occurs, {\it i.e.} second-class constraints arise. While Barros and Sa demonstrated \cite{BS} that these second-class constraints can be solved, such that only first class ones remain, Alexandrov provided us with a covariant formulation in which $\gamma$ does not enter the area spectrum\cite{A03}. Therefore, the development of a formulation, in which the Lorentz frame is not fixed, can provide a deep insight towards the understanding of gravitational quantum features.\\ 
In this work, we focus our attention on the role of the Lorentz symmetry after a canonical quantization of gravity in a second-order 4-bein formulation. We outline that if one solves classically constraints associated with boost symmetry, a parametric dependence of the wave functions on the reference frame cannot be avoided. But a unitary operator connecting states in different frames can be defined, such that the full Lorentz symmetry is implemented into the quantum framework. Then we perform the canonical quantization of all the classical constraints. By substituting the quantum boost constraints into the rotational ones, we get a similar picture to the previous case, but here the wave functional depends no longer parametrically on the Lorentz frame and it evolves through different values of the boost parameters. In this full quantization scheme, a natural operator representing the displacement of the boost parameters arises in a unitary form. This fact supports the idea of a gauge invariant dynamics, preserved by the quantization procedure.\\ 
The organization of the manuscript is as follows: in section 2 we describe the geometric interpretation of the configuration variables and develop a Hamiltonian formulation of General Relativity in terms of them. The algebra of constraints is analyzed and its first-class character is recognized. In section 3, at first we classically solve the constraints associated with the boost symmetry, demonstrating that transformations between Lorentz frames are implemented by a unitary operator. Hence we sketch properties of the quantum theory without any gauge fixing. Finally, in section 4 concluding remarks are provided.

\section{Geometric Structure and Hamiltonian formulation}
Our aim is to quantize geometric degrees of freedom in a canonical way. In particular, the configuration variables of our approach will be a set of 3-bein vectors, that, unlike standard treatments, are not restricted onto spatial hypersurfaces.\\ 
Let us consider an hyperbolic space-time manifold $V$ endowed with a metric $g_{\mu\nu}$ and a $3+1$ representation $V\rightarrow\Sigma\otimes R$, being $\Sigma$ spatial 3-hypersurfaces with internal coordinates $x^i\hspace{0.2cm}(i=1,2,3)$ and $t$ the coordinate on the real time-like axis. We perform a canonical quantization of 4-bein variables, but we want to avoid the usual time-gauge condition, {\it i.e.} the choice of the 3-bein $e_a,\hspace{0.2cm}(a=1,2,3)$ as contained into spatial hypersurfaces. In this respect, we introduce the following 4-bein 1-forms  
\begin{eqnarray}
e^0=Ndt+\chi_a E^a_idx^i\qquad e^a=E^a_iN^idt+E^a_idx^i,
\end{eqnarray}  
which define a generic Lorentz frame and the time-gauge is restored as soon as functions $\chi_a$ are set vanishing.\\ 
In view of giving a physical interpretation to $\chi_a$, we note that if we perform a local Lorentz transformation $\Lambda^A_{\phantom1B}$ on the tangent space (to set the 3-bein on $\Sigma$) the condition $\chi_a=-\Lambda^0_{\phantom1a}/\Lambda^0_{\phantom10}$ must stand. This fact leads us to identify $\chi_a$ with the velocity components of the $e^A$ frame with respect to one at rest, {\it i.e.} adapted to the spatial splitting. Moreover, since we are working with units $c=1$, the condition $\chi^2=\delta^{ab}\chi_a\chi_b<1$ must stand.\\   
The new expressions for the lapse function $\tilde{N}$, for the shift vector $\tilde{N}^i$ and for the 3-geometry $h_{ij}$  are obtained by the condition $e^A$ to be a 4-bein, {\it i.e.} $g_{\mu\nu}=\eta_{AB}e^A_\mu e^B_\nu$ (being $\eta_{AB}=diag\{-1;1;1;1\}$), and they turn out to be as follows     
\begin{eqnarray}
\tilde{N}=\frac{1}{\sqrt{1-\chi^2}}(N-N^iE_i^a\chi_a)\qquad\tilde{N}^i=N^i+\frac{E^c_l\chi_cN^l-N}{1-\chi^2}E^i_a\chi^a\quad\chi^a=\chi_b\delta^{ab}\nonumber\\
h_{ij}=E^a_iE^b_j(\delta_{ab}-\chi_a\chi_b)\label{3metr}.
\end{eqnarray}
The 3-bein vectors associated with $h_{ij}$ can be expressed in terms of $E^a_i$, i.e.  
\begin{equation}
E'^a_i=E^b_i(\delta^a_b-\alpha\chi^a\chi_b)\qquad\alpha=\frac{1-\sqrt{1-\chi^2}}{\chi^2}\label{Eprimo}
\end{equation}
and the last relation, just like the expression (\ref{3metr}), stresses how the dynamics of the spatial metric without the time-gauge condition is described by $E^a_i$ and $\chi_a$ variables, both.\\
As well-known, the canonical splitting of the Einstein-Hilbert action provides the Lagrangian density
\begin{equation}
\Lambda=\frac{1}{16\pi G}\tilde{N}\sqrt{h}(K^2-K_{ij}K^{ij}+{}^3\!R)
\end{equation}
$K_{ij}$ being the extrinsic curvature associated with $\Sigma$, {\it i.e.} $K_{ij}=\frac{1}{2\tilde{N}}(D_i\tilde{N}_j+D_j\tilde{N}_i-\partial_th_{ij})$, while ${}^{3}\!R$ is the scalar curvature of the 3-space.\\
In this formulation, $\tilde{N}$, $\tilde{N}^i$, $E^a_i$ and $\chi_a$ can be taken as configuration variables and their  conjugated momenta read as  
\begin{eqnarray}
\pi_{\tilde{N}}=0\qquad\pi_i=0\\
\pi^i_a=\frac{1}{8\pi G}\sqrt{h}[K^{ij}E_j^b(\delta_{ab}-\chi_a\chi_b)-KE_a^i]\\
\pi^a=\frac{1}{8\pi G}\sqrt{h}\bigg(\frac{\chi^a}{1-\chi^2}K-K^{ij}E^a_iE^b_j\chi_b\bigg),
\end{eqnarray}
respectively.
By virtue of the relation
\begin{equation}
K^i_j=\frac{8\pi G}{\sqrt{h}}\bigg(\pi^i_aE^a_j-\frac{1}{2}\delta^i_jE^b_l\pi^l_b\bigg)
\end{equation}
the equation below stands 
\begin{equation}
\pi^i_a\partial_tE^a_i+\pi^a\partial_t\chi_a=\frac{\sqrt{h}}{16\pi G}[-K h^{ij}\partial_t h_{ij}+K^{ij}\partial_th_{ij}].
\end{equation}
This way, one obtains
\begin{equation}
\pi^i_a\partial_tE^a_i+\pi^a\partial_t\chi_a-\Lambda=\tilde{N}'H+\tilde{N}^iH_i+\lambda^{\tilde{N}}\pi_{\tilde{N}}+\lambda^i\pi_i
\end{equation}
being $\tilde{N}'=\sqrt{h}\tilde{N}$, while $H$ and $H_i$ can be rewritten as
\begin{eqnarray}
H=\pi^i_a\pi^j_b\bigg(\frac{1}{2}E^a_iE^b_j-E^b_iE^a_j\bigg)+h{}^{3}\!R\label{H}\\
H_i=D_j(\pi^j_aE^a_i),\label{Hi}
\end{eqnarray}
with $D_i$ the covariant derivative built up from $h_{ij}$.\\
Moreover, phase space variables are not independent, but they are subjected to the following constraints
\begin{eqnarray}
\pi_{\tilde{N}}=0\qquad\pi_i=0\label{piNNi}\\
\Phi^a=\pi^a-\pi^b\chi_b\chi^a+\delta^{ab}\pi^i_b\chi_cE^c_i=0\label{boost}\\
\Phi_{ab}=\pi^c\delta_{c[a}\chi_{b]}-\delta_{c[a}\pi^i_{b]}E^c_i=0.\label{rot}
\end{eqnarray}
which are imposed by virtue of Lagrangian multipliers $\lambda^{\tilde{N}}$, $\lambda^i$, $\lambda_a$ and $\lambda^{ab}=-\lambda^{ba}$. Finally, the Hamiltonian density turns out to be
\begin{equation}
\mathcal{H}=\tilde{N}'H+\tilde{N}^iH_i+\lambda^{\tilde{N}}\pi_{\tilde{N}}+\lambda^i\pi_i+\lambda^{ab}\Phi_{ab}+\lambda_a\Phi^a.
\end{equation}
We want to stress that in the time gauge ($\chi_a=0$), conditions $\Phi^a=0$ do not arise.\\

\subsection{Dirac algebra of the constraints}
Let us now discuss the form of these constraints: the simplest ones are the four standard conditions (\ref{piNNi}), which induce the vanishing behavior of the super-Hamiltonian and of the super-momentum, as secondary constraints, {\it i.e.}
\begin{eqnarray}
H=0\qquad H_i=0.
\end{eqnarray}
As well-known, they account for the invariance under time re-parametrization and spatial diffeomorphisms, respectively, and their Poisson brackets vanish on the constraints hypersurfaces.\\
Other constraints enforce the invariance under 4-bein Lorentz transformations: in fact we have 
\begin{eqnarray}
\{\Phi^a;e^0\}=E^a_idx^i\qquad\{\Phi^a;e^c\}=\delta^{ac}E^d_i\chi_dN^idt+\delta^{ac}E^d_i\chi_ddx^i\\
\{\Phi_{ab};e^0\}=0\qquad\{\Phi_{ab};e^c\}=\delta_{d[a}\delta^c_{b]}e^d
\end{eqnarray}
and the above relations outline that $\Phi_{ab}$ and $\Phi^a$ act on the phase space as generators of rotations and boosts, modulo a time re-parametrization, respectively.\\   
If we introduce $\varphi^a=\epsilon^{abc}\Phi_{bc}$,  the boost-rotation algebra is clearly reproduced, in fact we have
\begin{eqnarray}
\{\Phi^a;\Phi^b\}=\epsilon^{ab}_{\phantom1\phantom2c}\varphi^c\quad\{\varphi^a;\varphi^b\}=-\epsilon^{ab}_{\phantom1\phantom2c}\varphi^c\quad\{\varphi^a;\Phi^b\}=-\epsilon^{ab}_{\phantom1\phantom2c}\Phi^c.\label{boost-rot}
\end{eqnarray} 
Since Lorentz transformations do not modify the 3-metric ($\{\Phi^a;h_{ij}\}=\{\Phi_{ab};h_{ij}\}=0$) and 
\begin{eqnarray}
\{\Phi^a;\pi_c^iE_j^c\}=\{\Phi_{ab};\pi_c^iE_j^c\}=0\\\{\Phi^a;\pi^i_c\pi^j_d\bigg(\frac{1}{2}E^c_iE^d_j-E^d_iE^c_j\bigg)\}=\{\Phi_{ab};\pi^i_c\pi^j_d\bigg(\frac{1}{2}E^c_iE^d_j-E^d_iE^c_j\bigg)\}=0
\end{eqnarray}
we find the following last relations which determine the algebra of constraints
\begin{equation}
\{\Phi^a;H\}=\{\Phi_{ab};H\}=0\qquad\{\Phi^a;H_i\}=\{\Phi_{ab};H_i\}=0.\label{br-hm}
\end{equation}
Therefore, conditions (\ref{boost-rot}) and (\ref{br-hm}) demonstrate that the set of constraints is of first class.\\
We want to stress that, being associated with first class constraints, the symmetry under boosts actually plays the role of a gauge symmetry and no second-class constraint arises, unlike the issues discussed by Alexandrov \cite{A00}.\\

\section{Quantization of the model} 
Let us provide a classical solution for boost constraints (\ref{boost}). One can solve it for $\pi^a$ (since they enter linearly in $\Phi^a$) getting the following expression 
\begin{equation}
\pi^a=-\bigg(\delta^{ab}+\frac{\chi^a\chi^b}{1-\chi^2}\bigg)\pi^i_b\chi_cE^c_i.
\end{equation}
Hence, we can fix the boost symmetry by giving functions $\chi_a=\bar{\chi}_a(t;x)$. In order to deal with a pure constrained Hamiltonian theory, we simplify the dynamics by choosing a Lorentz frame which moves with constant velocity, thus $\partial_t\bar{\chi}_a=0$. From Hamilton equations we have
\begin{equation} 
\partial_t \bar{\chi}_a=\lambda^b(\delta_{ab}-\bar{\chi}_a\bar{\chi}_b)+\lambda^{ab}\bar{\chi}_b=0
\end{equation}
which allows one to write $\lambda^a=-\lambda^{ab}\bar{\chi}_b$. The possibility to express the Lagrangian multipliers $\lambda^a$ in terms of those ones $\lambda^{ab}$ reflects how they become redundant, when the boost constraints are solved. Hence in this case we can rewrite the action as follows
\begin{equation}
S=-\frac{1}{16\pi G}\int[\pi^i_a\partial_tE^a_i+\pi_{\tilde{N}'}\partial_t\tilde{N}'+\pi_i\partial_t\tilde{N}^i-\tilde{N}'H^{\bar\chi}-\tilde{N}^iH^{\bar\chi}_i-\lambda^{ab}\Phi'_{ab}-\lambda^{\tilde{N}}\pi_{\tilde{N}}-\lambda^i\pi_i]dtd^3x,
\end{equation}
where
\begin{equation}
\Phi'_{ab}=\bar{\chi}_{[a}\pi^i_{b]}E^d_i\bar{\chi}_d-\delta_{c[a}\pi^i_{b]}E^c_i\label{rotcon}
\end{equation}
gives the new form of the constraints, while $H^{\bar\chi}$ and $H^{\bar\chi}_i$ are the super-Hamiltonian and the super-momentum with variables $\chi$ replaced by functions ${\bar\chi}$. This set of constraints is again first-class.\\
In this picture we have completely fixed the gauge associated with the boost symmetry, because $\bar{\chi}_a$ are three functions to be assigned explicitly together with the Cauchy data.\\ 
Nevertheless, we see how a dynamics is obtained, which differs from that one in which the time-gauge is imposed: this just because of a relic dependence on parameters $\bar{\chi}_a$. From a geometrical point of view, this issue is not surprising, since our configuration variables $E^a_i$  are no more 3-bein within spatial hypersurfaces, but they still remain variables which contribute to the 3-metric (indeed they are now projections of the 3-bein over the spatial hypersurfaces).\\
We emphasize that, by adopting the variables $\bar{\chi}_a(x^i)$ as new coordinates, we could not eliminate their effect on the dynamics, because the constraints contain such quantities free of spatial derivatives, too.\\ 
In view of the quantization, we now promote to operators $\tilde{N}$, $\tilde{N}^i$, $E^a_i$ and the corresponding  conjugated momenta, we replace Poisson brackets with commutators in a canonical way and hence we impose relic constraints on wave functionals $\psi=\psi_{\bar\chi} (\tilde{N},\tilde{N}^i,E^a_i)$.\\ 
In particular, conditions (\ref{piNNi}) are translated into $\frac{\delta}{\delta\tilde{N}}\psi=\frac{\delta}{\delta \tilde{N}^i}\psi=0$, thus $\psi$ does not depend on $\tilde{N}$ and $\tilde{N}^i$. Hence the super-momentum constraint reads as follows\footnote{We will not take into account of ordering questions, since they do not modify our conclusions.}
\begin{equation}
\hat{H}^{\bar\chi}_i\psi_{\bar\chi}(E)=iD_j\bigg(E^a_i\frac{\delta}{\delta E^a_j}\bigg)\psi_{\bar\chi}(E)=0\label{qsm}
\end{equation}
and it implies that wave functionals do not change for $E^a_i\rightarrow E^a_i-D_i\xi^jE^a_j$, being $\xi^i$ an arbitrary 3-vector. This means that $\psi$ depends on the classes $\{E^a_i\}$, built up by identifying $E^a_i$ related by the above transformation, {\it i.e.} infinitesimal 3-diffeomorphisms.\\ 
A further restriction for $\psi$ is provided by the rotational quantum constraints, whose form is as follows
\begin{equation}
\Phi'^{\bar\chi}_{ab}\psi_{\bar\chi}(E)=i\bigg[{\bar\chi}_{[a}\frac{\delta}{\delta E_i^{b]}}E^d_i{\bar\chi}_d-\delta_{c[a}\frac{\delta}{\delta E_i^{b]}}E^c_i\bigg]\psi_{\bar\chi}(E)=0:\label{qrotcon}  
\end{equation}
it outlines the relation existing, in this approach, between the wave-functional dependence on $E^a_i$ and the choice of the functions ${\bar\chi}_a(x^i)$.\\ 
Finally, the dynamics comes out from 
\begin{equation}
\hat{H}^{\bar\chi}\psi_{\bar\chi}(E)=\bigg[-\bigg(\frac{1}{2}E^a_iE^b_j-E^b_iE^a_j\bigg)\frac{\delta}{\delta E_i^a}\frac{\delta}{\delta E_j^b}+h{}^{3}\!R\bigg]\psi_{\bar\chi}(E)=0,\label{qsh}
\end{equation}
which clarifies how ${\bar\chi}_a$ do not disappear from the quantum description, but, being all the constraints dependent on ${\bar\chi}$, wave functionals contain ${\bar\chi}$ as labels.

\subsection{Transformation between $\chi$-sectors}

In order to investigate if the transformation between different ${\bar\chi}$-sectors can be implemented in a quantum setting, an operator connecting Hilbert spaces with different forms of $\bar{\chi}$ must be defined.\\ 
Let us now consider a wave functional $\psi_0$ in the time gauge: it is a solution of the following system of constraints
\begin{equation}
H^{0}\psi_0=0\quad H^{0}_i\psi_0=0\quad-\delta_{c[a}\pi^i_{b]}E^c_i\psi_0=0,
\end{equation}
$H^{0}$ and $H^{0}_i$ being the super-Hamiltonian and super-momentum built up from the metric tensor $h_{ij}=\delta_{ab}E^a_iE^b_j$, {\it i.e.} in the case $\bar{\chi}\equiv0$.\\
Taking into account the operator $U$ 
\begin{equation}
U_\epsilon=I-\frac{i}{4}\int\epsilon^a\epsilon_b(E^b_i\pi^i_a+\pi^i_aE^b_i)d^3x+O(\epsilon^4),\label{U}
\end{equation}
responsible for the transformation
\begin{equation}
U_\epsilon E^a_iU_\epsilon^{-1}=E^b_i(\delta^a_b-\frac{1}{2}\epsilon^a\epsilon_b)+O(\epsilon^4)={E'}^a_i+O(\epsilon^4)
\end{equation}
which maps the metric $h_{ij}$ from $\bar{\chi}=0$ to $\bar{\chi}_a=\epsilon_a\ll1$,
then, after same algebra, the state $\psi'=U_\epsilon\psi_0$ can be rewritten as 
\begin{equation}
\psi'(E)=\psi_0(E').
\end{equation}
The new state will satisfy
\begin{equation}
U_\epsilon H^0U_\epsilon^{-1}\psi'=0\quad U_\epsilon H^0_iU_\epsilon^{-1}\psi'=0\quad U_\epsilon(-\delta_{c[a}\pi^i_{b]}E^c_i)U^{-1}_\epsilon\psi'=0.
\end{equation}
Since we have
\begin{equation} 
U_\epsilon
E^a_i\pi^j_aU_\epsilon^{-1}=E^a_i\pi^j_a+O(\epsilon^4), 
\end{equation}
$H^{0}$ and $H^{0}_i$ are translated in $H^\epsilon$ and $H^\epsilon_i$ up to the $\epsilon^2$ order.\\ 
Moreover, rotational constraints becomes
\begin{eqnarray}
-\bigg[\delta_{c[a}\pi^i_{b]}E^c_i+\frac{1}{2}\delta_{c[a}\epsilon_{b]}\epsilon^dE^c_i\pi^i_d-\frac{1}{2}\epsilon_d\epsilon_{[a}\pi^i_{b]}E^d_i\chi_d+O(\epsilon^4)\bigg]\psi'=0\label{Ucon}
\end{eqnarray}
and, starting from this condition, the expression $\epsilon_dE^d_i\pi^i_b$ can be calculated, multiplying it times $\epsilon^a$ and retaining the leading orders in $\epsilon_a$. Thus by substituting this result into (\ref{Ucon}), the constraints $\Phi'_{ab}$ (\ref{qrotcon}) come out for ${\bar\chi_a=\epsilon_a}$.\\  
Hence the operator $U_\epsilon$ implements the mapping of physical states corresponding to ${\bar\chi}=0$ and ${\bar\chi}=\epsilon$. For this reason we will indicate $\psi'$ with $\psi_\epsilon$. We emphasize that, since $U^{-1}=U^\dag$, then the transformation between a frame at rest and one moving with respect to $\Sigma$ can be implemented by a unitary operator.\\
As can be checked explicitly from the theory of constrained systems \cite{HT}, $U_\epsilon$ is given by the exponential of the boost constraint $\exp(i\int d^3 x \epsilon_a \Phi^a)$. In fact in the relation (\ref{U}) we have part of the quadratic term in the $\bar{\chi}$-expansion of this operator, and these two transformations coincide, as far as one recognizes that for $\bar{\chi}_a=0$ $\pi^a$ and $\chi_a$ are not configuration variables anymore. This correspondence allows to reproduce the operator $U_\epsilon$ for any value of $\bar{\chi}_a$ and at all orders in a perturbative expansion.

\subsection{Quantization without gauge fixing}
A different approach with respect to that of the previous section is one in which $\chi_a$ are not fixed.\\ 
In this respect we promote also $\chi_a$ and their conjugate momenta to operators on a Hilbert space. Hence we impose the full set of constraints on a wave functional $\psi=\psi(\tilde{N},\tilde{N}^i,E^a_i,\chi_a)$, such that solutions provide us with physical states. The independence of wave functionals from $\tilde{N}$ and $\tilde{N}^i$ is again recovered. The super-momentum (\ref{qsm}) and the super-Hamiltonian (\ref{qsh}) are formally not modified, despite the fact that $\chi_a$ is now a real quantum variable.\\ 
Otherwise, rotational constraints and boost ones are
\begin{eqnarray}
\hat{\Phi}_{ab}\psi(E,\chi)=i\bigg(\frac{\delta}{\delta\chi_c}\delta_{c[a}\chi_{b]}-\delta_{c[a}\frac{\delta}{\delta E_i^{b]}}E^c_i\bigg)\psi(E,\chi)=0\\
\hat{\Phi}^a\psi(E,\chi)=i\bigg(\frac{\delta}{\delta\chi_a}-\frac{\delta}{\delta\chi_b}\chi_b\chi^a+\delta^{ab}\frac{\delta}{\delta E_i^b}\chi_cE^c_i\bigg)\psi(E,\chi)=0.
\end{eqnarray}
Substituting the boost constraints into the rotational ones, we easily recognize that the latter retain formally the same expression as $\Phi'_{ab}$ (\ref{qrotcon}). But the presence of the boost constraints give an ``evolutionary'' character of the wave-functional on the $\chi$-variables. In this respect, we remark the non-vanishing character of the conjugate momenta $\pi^a$, when acting on physical states.\\ 
In this framework, one cannot speak of transformations between $\chi$-sectors, being $\chi_a$ operators and the Hilbert space is necessary a unique one. Nevertheless, one can formally implement translations on $\chi_a$ by using their conjugated variables $\pi^a$ as generators. This transformation $\hat{T}=I-\epsilon_a(x,t)\frac{\delta}{\delta \chi_a};\hspace{0.2cm}\epsilon_a\ll1$ turns out to be unitary.\\ 
Therefore, we expect that the Lorentz symmetry is not affected by the quantization as soon as also $\chi_a$ are quantized.

\section{Concluding remarks}
We have performed the canonical quantization of General Relativity in a 4-bein formulation, by dropping one of the standard assumption, {\it i.e.} the time-gauge condition. This way we are dealing with a Lorentz frame moving with respect to spatial hypersurfaces, so that we have three additional Lagrangian variables, $\chi_a$, giving the velocity components of such motion. As a consequence of the boost invariance, three new constraints arise, whose algebra results to be of first class. We have classically solved these constraints and we found that the $\chi_a$'s do not disappear from the dynamics, but they play a parametric role. Furthermore, we have canonically quantized the system and we recovered an infinitesimal unitary operator, mapping physical states in the time gauge into the corresponding for $\chi_a\neq0$. Moreover such a kind of operators, realizing $\chi$-translations, can be defined also in the case in which $\chi_a$ are quantized, too.\\ 
These issues indicate that the invariance under boost transformations is preserved on a quantum level, {\it i.e.} that scalar products are not modified in different $\chi$-sectors. This provide us with an explanation for the use of the time-gauge condition, because any other choice for the Lorentz frame gives the same expectation values for observables.\\
To physically characterize spatial hypersurfaces, a matter field can be introduced, as it will be illustrated in \cite{CMprox}.\\


\begin{thebibliography}{99}

\bibitem{Pol}
J. Polcinsky, {\it String Theory}, Cambridge University Press, (1998).

\bibitem{Rov}
C. Rovelli, {\it Quantum Gravity}, Cambridge University Press, (2004).

\bibitem{Ba95}
J. F. Barbero, {\it Phys. Rev.}, {\bf D51}, 10, (1995), 5507. 

\bibitem{ABCK}
A. Ashtekar, J. Baez, A. Corichi, K. Krasnov, {\it Phys. Rev. Lett.}, {\bf 80}, (1998), 904.

\bibitem{Kr}
 I.B. Khriplovich, {\it J. Exp. Theor. Phys.}, {\bf 100}, (2005), 1075; {\it Zh. Eksp. Teor. Fiz.}, {\bf 100}, (2005) 1223.

\bibitem{GM}
L. J Garay, G. A. M. Marugan, {\it Class. Quantum Grav.}, {\bf 20}, 8, (2003), L115. 

\bibitem{A00}
S. Alexandrov, {\it Class. Quant. Grav.}, {\bf 17}, (2000), 4255.

\bibitem{BS}
N. Barros e Sa, {\it Int. J. Mod. Phys.}, {\bf D10}, (2001), 261.

\bibitem{A03}
S. Alexandrov, E. R. Livine, {\it Phys. Rev.}, {\bf D67}, (2003), 044009.


\bibitem{HT}
M. Henneaux, C. Teitelboim, \emph{Quantization of Gauge Systems}, Princeton University Press, (1994).

\bibitem{CMprox}
F. Cianfrani, G. Montani, ``The role of matter fields as observers in Quantum Gravity without time gauge'', in preparation.


\end{thebibliography}
\end{document}